# Analyzing Noise Models and Advanced Filtering Algorithms for Image Enhancement

**Sahil Ali Akbar**
Department of Computational Intelligence, SRM Institute of Science and Technology, Kattankulathur, India - 603203
E-mail: sa4612@srmist.edu.in
ORCID iD: https://orcid.org/0009-0004-6655-2112

**Ananya Verma**
Department of Computing Technologies, SRM Institute of Science and Technology, Kattankulathur, India - 603203
Email: av7601@srmist.edu.in
ORCID iD: https://orcid.org/0009-0009-8652-2785

**Abstract:** Noise, an unwanted component in an image, can be the reason for the degradation of Image at the time of transmission or capturing. Noise reduction from images is still a challenging task. Digital Image Processing is a component of Digital signal processing. A wide variety of algorithms can be used in image processing to apply to an image or an input dataset and obtain important outcomes. In image processing research, removing noise from images before further analysis is essential. Post-noise removal of images improves clarity, enabling better interpretation and analysis across medical imaging, satellite imagery, and radar applications. While numerous algorithms exist, each comes with its own assumptions, strengths, and limitations. The paper aims to evaluate the effectiveness of different filtering techniques on images with eight types of noise. It evaluates methodologies like Wiener, Median, Gaussian, Mean, Low pass, High pass, Laplacian and bilateral filtering, using the performance metric Peak signal to noise ratio. It shows us the impact of different filters on noise models by applying a variety of filters to various kinds of noise. Additionally, it also assists us in determining which filtering strategy is most appropriate for a certain noise model based on the circumstances.

**Index Terms:** Noise Removal, PSNR, Digital Image Processing, Digital Signal Processing, Image filters, Image Noise.

## 1. Introduction

Noise commonly persists in the majority of images, impacting their quality and clarity. The process of image denoising involves the removal of unwanted disturbances from the image, which simplifies the task of interpreting and analyzing the image. A crucial characteristic of an effective image denoising model is its ability to thoroughly eliminate noise while also preserving the sharpness of edges within the image [1]. Filters play a pivotal role in advancing visual enhancement and interpretation, thereby establishing a fundamental basis for image segmentation. Additionally, this contributes to various processes such as interpolation and resampling, further enhancing the overall image quality and accuracy [2]. The choice of filter utilized during analysis is contingent upon the characteristics of the image data. Two primary categories of filters are linear and non-linear filters [3].

## 2. Problem Statement

To develop an effective image denoising model that can robustly eliminate noise while preserving the sharpness of edges, thereby enhancing the overall image quality and facilitating accurate interpretation and analysis.

## 3. Problem Solution

Let Fig. 1 be the image used for applying noise and then different filters will be applied to evaluate the effectiveness of that filter to that particular type of noise.

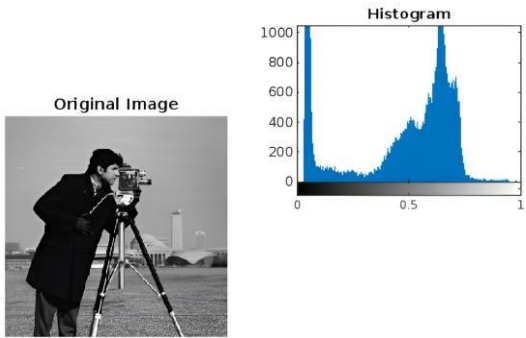

Fig. 1. Original Image used for evaluation

Fig. 2 and Fig. 3 shows the flowcharts which will be used for evaluating the methodologies.

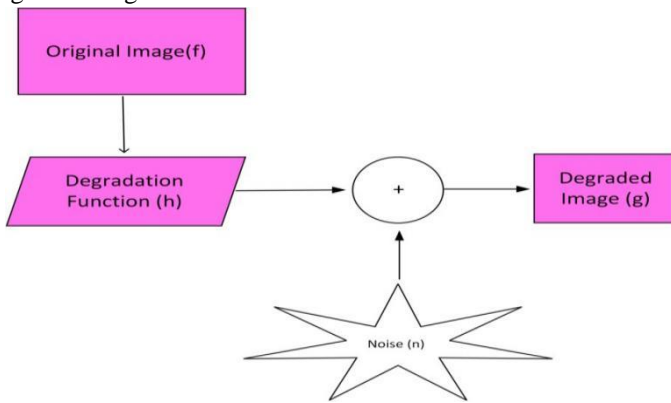

Fig. 2. The Diagram of the flowchart for adding noise to the Image

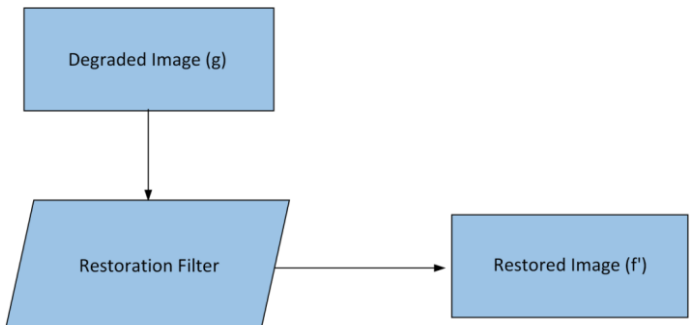

Fig. 3. The Diagram of the flowchart for applying the Noise filters for the noisy Image.

## 4. Types of Noise in Images

### 4.1 Gaussian Noise (Amplifier Noise)
As Gaussian noise occurs in amplifiers or detectors, we can alternatively call it electronic noise [4]. Natural sources of Gaussian noise include the discrete radiation nature of warm objects and the thermal vibrations of atoms [5]. In digital images, Gaussian noise typically disrupts gray values, leading to the development of a Gaussian noise model which can be characterized by its Probability Density Function (PDF) or a normalized histogram based on gray values.

The PDF of Gaussian Random Variable is given by

$$F(g) = \sqrt{\frac{1}{2\pi\sigma^2}} \frac{-(g-\mu)^2}{2\sigma^2}$$

F(g) = Gaussian distribution noise in image
σ = Standard Deviation
μ = Mean Value

The normalized Gaussian noise curve shows that this Gaussian noise model has a bell-shaped curve because of its equal randomness. Between 70% and 90% of the pixel values in the degraded image fall within the range of μ ± σ, according to the PDF of this noise model [4]. In the spectral domain, the normalized histogram's shape is constant, as shown in Fig. 4, which shows the PDF of Gaussian noise [4].

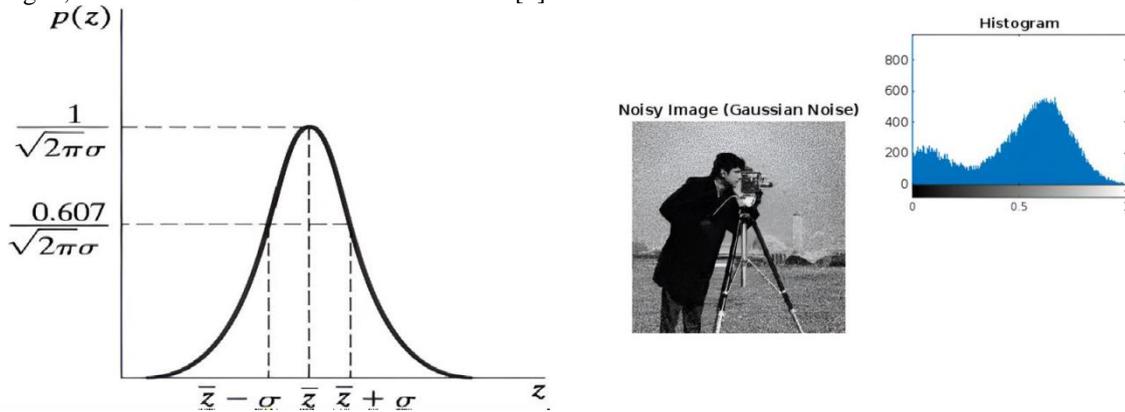

Fig. 4. PDF of Gaussian Noise (left side), Image after applying the noise along with Histogram (right side)

### 4.2 Salt and Pepper Noise (Impulse Noise)

Salt-and-pepper noise refers to the sporadic occurrence of white and black pixels occasionally observed in images. Both morphological and median filtering are frequently used techniques for lowering this kind of noise [6, 7]. Due to information loss and unwanted visual artifacts, noise negatively impacts image quality; one of the main causes of this degradation is salt-and-pepper noise. Other names for this kind of noise are spike noise, random noise, and impulse noise. Salt-and-pepper noise can be caused by a number of things, including hardware malfunctions, software bugs, or flaws in the camera sensor during the taking or sending of images. In contrast to other forms of noise, salt-and-pepper noise only affects a portion of the image's pixels, leaving the rest intact [8].

Salt and pepper noise can have values of either minimum (0) or maximum (255). For pepper noise, the values are close to 0, and for salt noise, they are close to 255. The intensity values of salt-and-pepper noise typically fall at either the minimum (0) or maximum (255) levels.

$$\eta(x,y) = \begin{cases} 0, pepper\ noise \\ 255, salt\ noise \end{cases}$$

The PDF of Salt and Pepper Noise is given by

$$p(z) = \begin{cases} p_a & for\ z = a \\ p_b & for\ z = b \\ 0 & otherwise \end{cases}$$

Where z is the random variable representing noise
p(z) = probability density function

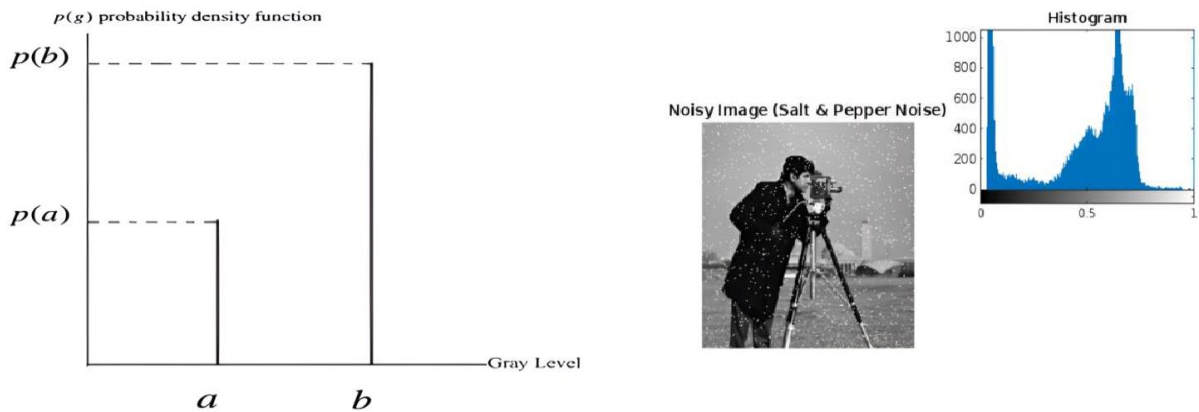

Fig. 5. PDF of Salt and Pepper Noise (left side), Image after applying Salt and Pepper Noise along with Histogram (right side)

The image shows two spikes in the bright region (region a) and dark region (region b) of salt-and-pepper noise, with the probability density function values at their minimum and maximum respectively [9]. Fig. 5 provides an illustration of salt-and-pepper noise.

**4.3 Speckle Noise**

Speckle noise, or multiplicative noise, is a phenomenon resulting due to the interaction between laser light and the roughness of the PSi surface [10]. The resulting signal is a product of the original signal and the speckle noise [11]. We can represent the multiplicative noise model as:

I(i,j)=M(i,j)×N(i,j)

The text describes an image pixel model where (i, j) represents a distorted pixel, M(i, j) represents the noiseless pixel, and N(i, j) represents the speckle noise signal [12]. The speckle noise signal's probability density function follows a gamma distribution, as illustrated in Fig. 6.

The PDF of Speckle Noise is given by

$$F(g) = \frac{g^{\alpha-1}}{(\alpha-1)!\,\alpha^{\alpha}} e^{\frac{-g}{\alpha}}$$

Where $\alpha^2$ is variance and g is the gray level.

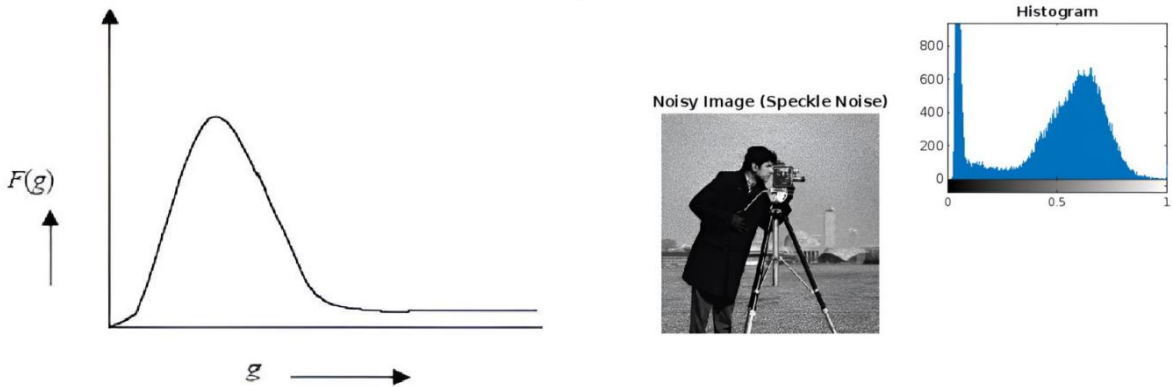

Fig. 6. PDF of Speckle Noise (left side), Image after applying Speckle Noise along with Histogram (right side)

**4.4 Poisson Noise (Photon Noise)**

This phenomenon is produced by the statistical properties of electromagnetic waves, which include x-rays, gamma rays, and visible light [4]. When used in medical imaging configurations, x-rays and gamma ray's emissions are directed into the patient's body from their respective sources [4]. These sources manifest random fluctuations in photon emission rates, thereby yielding images imbued with spatial and temporal irregularities. This form of noise is also known as quantum (photon) noise or shot noise [4]. This noise follows Poisson distribution and is given by

$$P\big((f_{(pi)} = k\big) = \frac{\lambda^k i^{e-\lambda}}{k!}$$

Where λ is the mean
K = 0,1,2,….

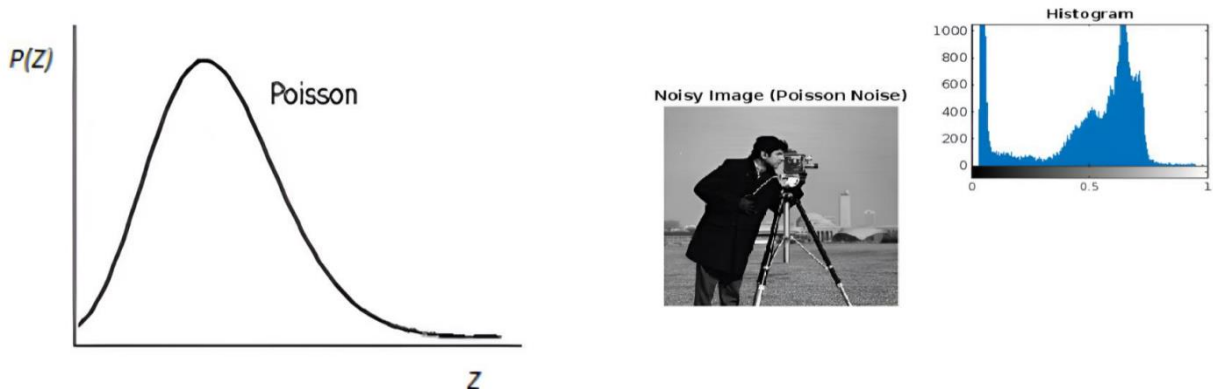

Fig. 7. PDF of Poisson Noise (left side), Image after applying Poisson Noise along with Histogram (right side)

**4.5 Periodic Noise (Sinusoidal Noise)**

    Periodic noise, a unique type of noise, is a challenge in image acquisition due to electrical and electromechanical interfaces, which cannot be effectively eliminated in the spatial domain. Spatial domain masks are generally ineffective for removing periodic noise, making it virtually impossible to address this issue through traditional means [13].

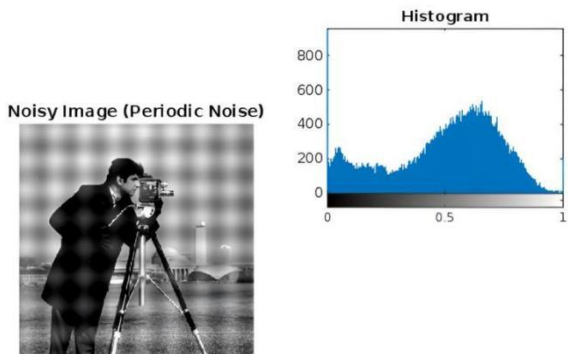

Fig. 8. Image after applying Periodic Noise along with Histogram

**4.6 Erlang Noise (Gamma Noise)**

    Gamma noise is a common feature in laser-based images. It follows the Gamma distribution and is depicted in the Probability Density Function (PDF) in Fig. 9.

PDF of Erlang Noise is given by

$$P(z) = \begin{cases} \dfrac{a^b z^{b-1}}{(b-1)!} e^{-az}, & for\ z \geq 0 \\ 0, & otherwise \end{cases}$$

Where the parameter a>0, positive integer is b.

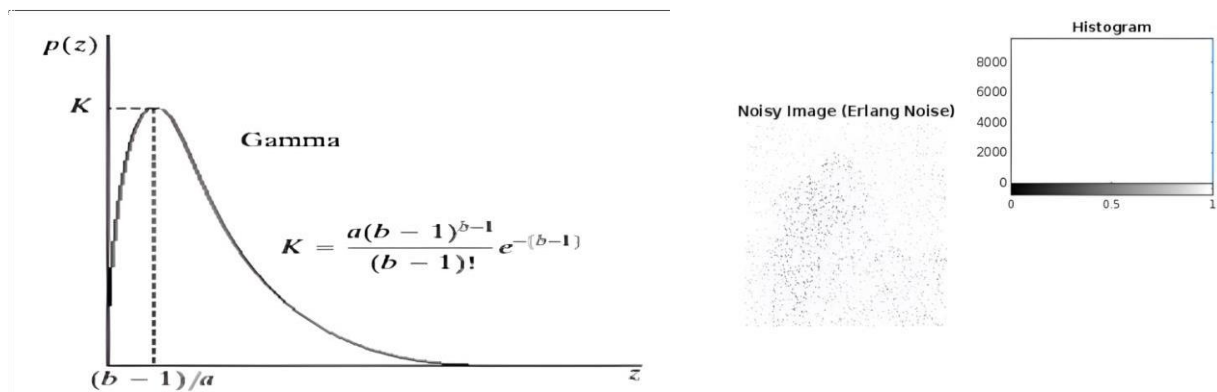

Fig. 9. PDF of Gamma Noise (left side), Image after applying Gamma Noise along with Histogram (right side)

### 4.7 Exponential Noise
The exponential distribution serves as the probability distribution that characterizes the timing between events in a Poisson process. It is a case of Exponential Noise where b = 1.
PDF of Exponential Noise is given by

$$P(z) = \begin{cases} ae^{-az}, & for\ z \geq 0 \\ 0, & otherwise \end{cases}$$

Where the paramtere a>0.

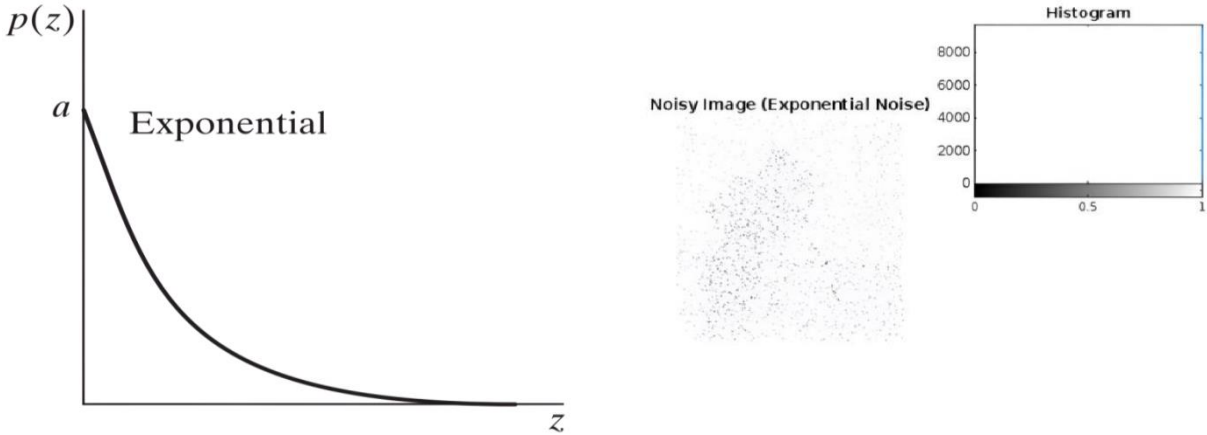

Fig. 10. PDF of Exponential Noise (left side), Image after applying Exponential Noise along with Histogram (right side)

### 4.8 Rayleigh Noise
Rayleigh noise occurs frequently in radar range images. In this context, the probability density function is defined as specified [14]. Unlike the Gaussian distribution, the Rayleigh distribution is characterized by its lack of symmetry [13]. The PDF of Rayleigh Noise is given by

$$P(z) = \begin{cases} \frac{2}{b}(g-a)e^{\frac{-(g-a)^2}{b}}, & for\ g \geq a \\ 0, & for\ g < a \end{cases}$$

Where , mean μ=a+$\sqrt{\frac{\pi b}{4}}$ and variance $\sigma^2 = \frac{b(4-\pi)}{4}$

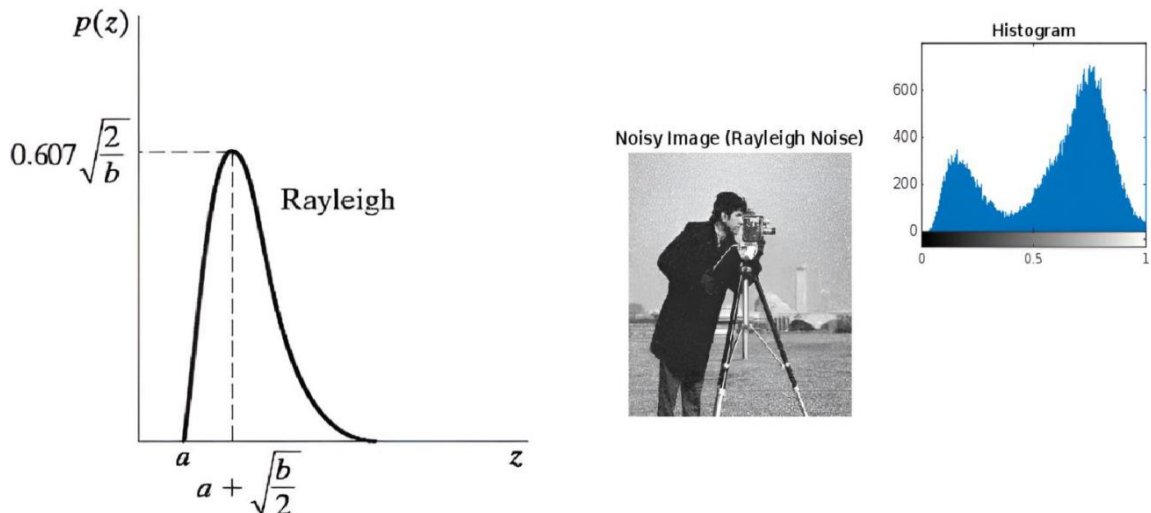

Fig. 11. PDF of Rayleigh Noise (left side), Image after applying Rayleigh Noise (right side)

## 5. Filtering Techniques

**5.1 Median Filter**

Median filter is a nonlinear spatial digital filtering technique. Its main use is for noise reduction and smoothing. It's very effective at preserving the image's edges while simultaneously removing the noise. The median filter works best for salt and pepper noise although we can apply it to other noise models as well. Its effectiveness will vary depending on the type of noise. The median filter analyzes every pixel in a given neighborhood, collects intensity values, and chooses the median value to remove impulsive noise from an image. The process produces a denoised image while maintaining image features and edges without unnecessary blurring. The formula for median filter is

$$\text{Mean filter}(x_1, x_2, \ldots \ldots x_N) = \frac{1}{N}\sum_{i=1}^{N} x_i$$

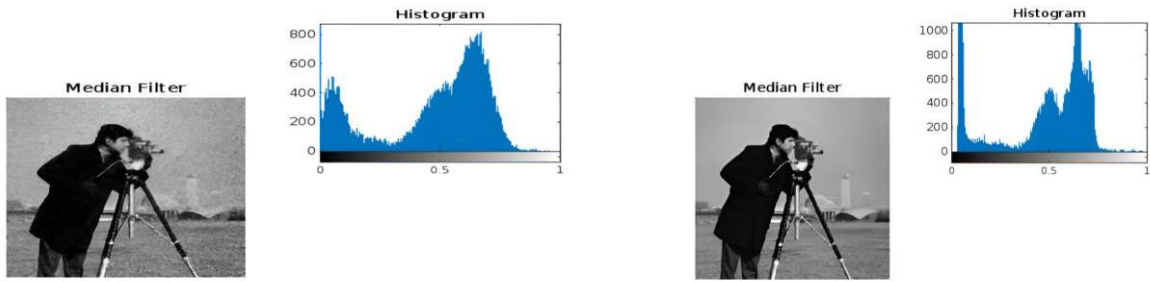

Fig. 12. Median Filter applied on Image with Gaussian Noise along with histogram (left side), Median Filter applied on Image with Salt and Pepper Noise along with histogram (right side)

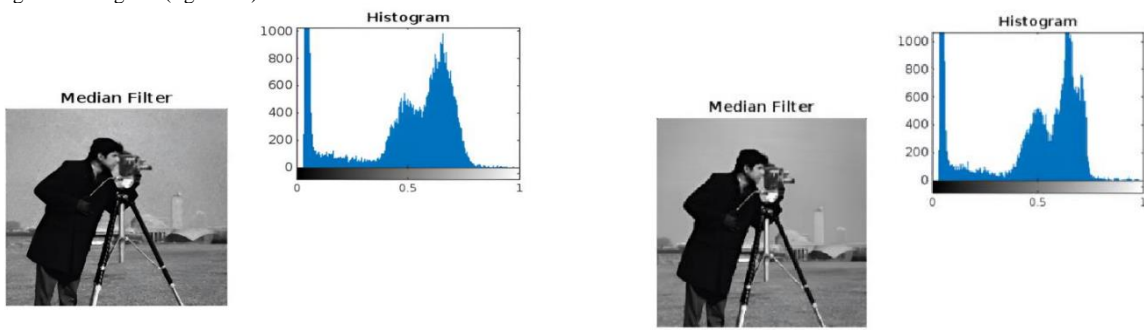

Fig. 13. Median Filter applied on Image with Speckle Noise along with histogram (left side), Median Filter applied on Image with Poisson Noise along with histogram (right side)

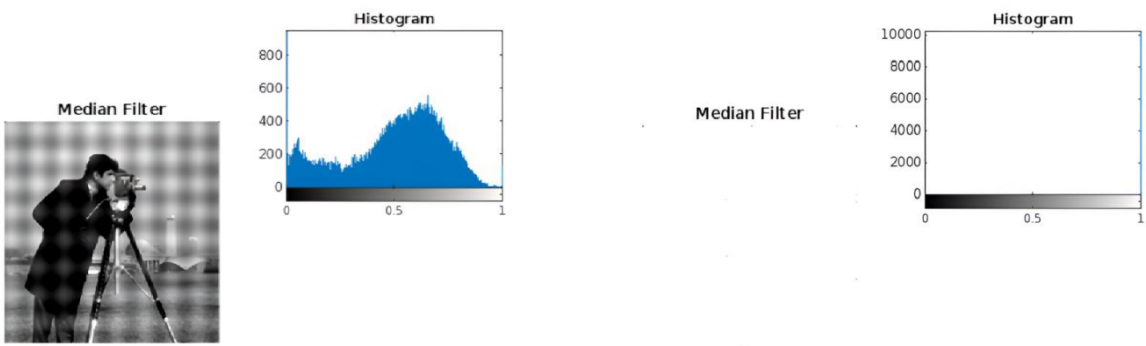

Fig. 14. Median Filter applied on Image with Periodic Noise along with histogram (left side), Median Filter applied on Image with Erlang Noise along with histogram (right side)

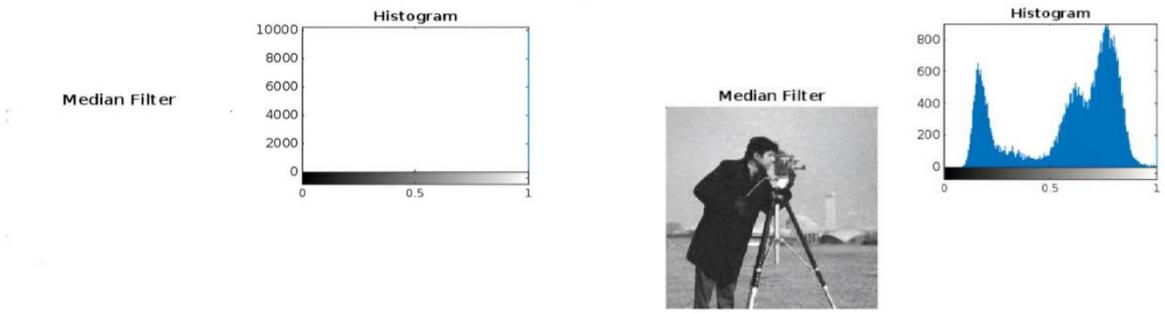

Fig. 15. Median Filter applied on Image with Exponential Noise along with histogram (left side), Median Filter applied on Image with Rayleigh Noise along with histogram (right side)

**5.2 Mean Filter**

A mean filter is a linear spatial filter in digital image processing. It is used for noise reduction and smoothing. It is more effective at reducing less severe noise in images. It is well suited for Gaussian and uniform noise but can be applied to other noise models as well depending on their characteristics. One of the drawbacks of the mean filter is edge blurring. It is not able to preserve the fine details in the image. By averaging pixel values within a short window, the mean filter produces a smoothing effect that lowers image noise. However, utilizing a big window size can make the edges blurry. The formula for mean filter is given by

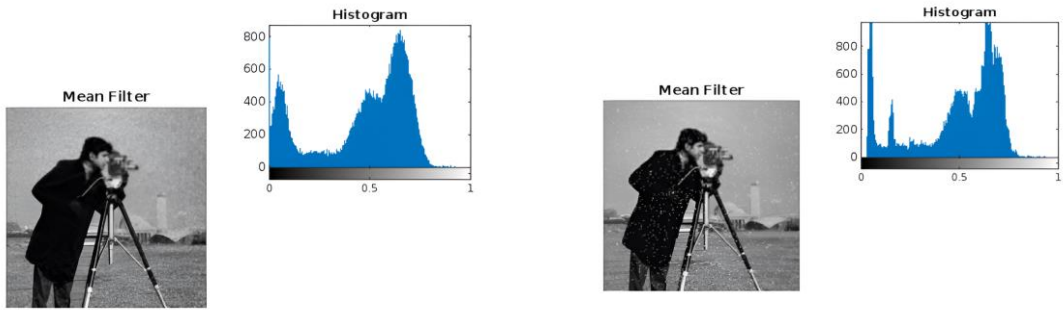

Fig. 16. Mean Filter applied on Image with Gaussian Noise along with histogram (left side), Mean Filter applied on Image with Salt and Pepper Noise along with histogram (right side)

$$\text{Mean filter}(x_1, x_2, \ldots \ldots x_N) = \frac{1}{N}\sum_{i=1}^{N} x_i$$

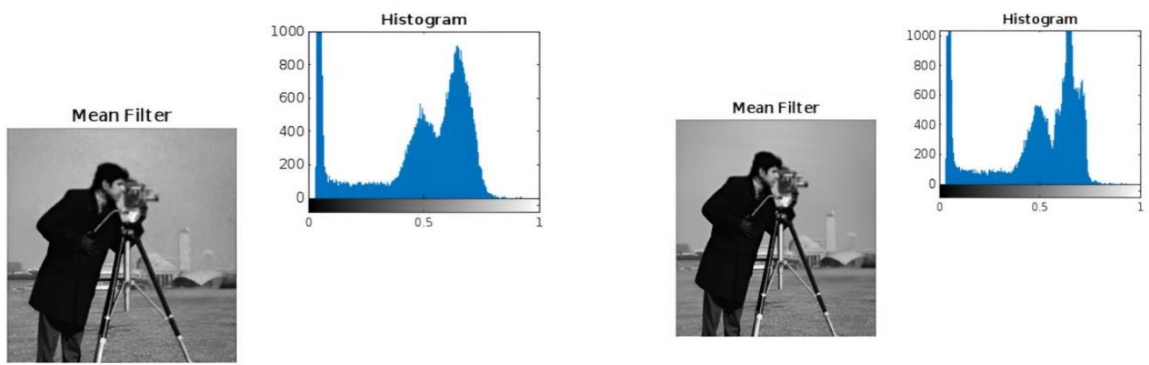

Fig. 17. Mean Filter applied on Image with Speckle Noise along with histogram (left side), Mean Filter applied on Image with Poisson Noise along with histogram (right side)

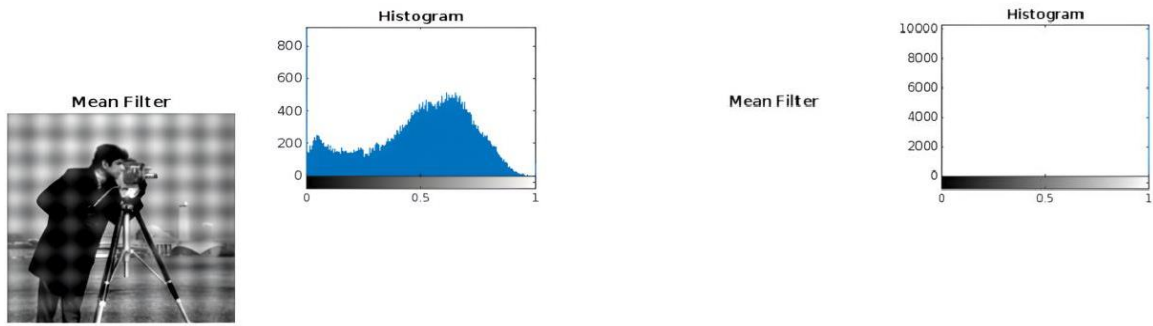

Fig. 18. Mean Filter applied on Image with Periodic Noise along with histogram (left side), Mean Filter applied on Image with Erlang Noise along with histogram (right side)

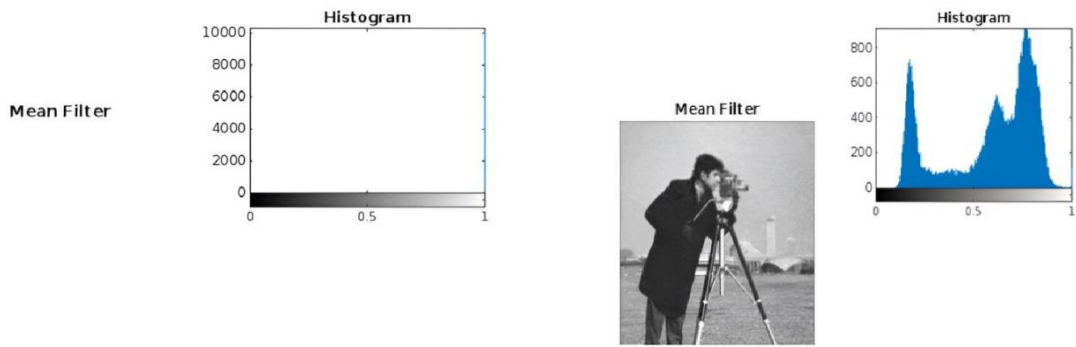

Fig. 19. Mean Filter applied on Image with Exponential Noise along with histogram (left side), Mean Filter applied on Image with Rayleigh Noise along with histogram (right side)

### 5.3 Wiener Filter

When the statistical characteristics of the signal and noise are well-understood or accurately predicted, the Wiener filter proves to be an effective tool for image denoising and restoration. It is essentially a statistical filter that has a practical implementation in both the spatial and frequency domains. It works best for the noise models that are additive and stationary like Gaussian noise and not so well for salt and pepper noise. Considering both the statistical characteristics of the noise and the noisy signal, the Wiener filter predicts the original signal from noisy observations in the frequency domain. Power spectral density (PSD) and frequency-dependent weighting are used to reduce noise while keeping the main parts of the original signal. The formula for Wiener filter is

$$W(f_1, f_2) = \frac{H*(f_1, f_2) S_{xx}(f_1 f_2)}{(|H(f_1, f_2)|)^2 S_{xx}(f_1, f_2) + S_{nn}(f_1, f_2)}$$

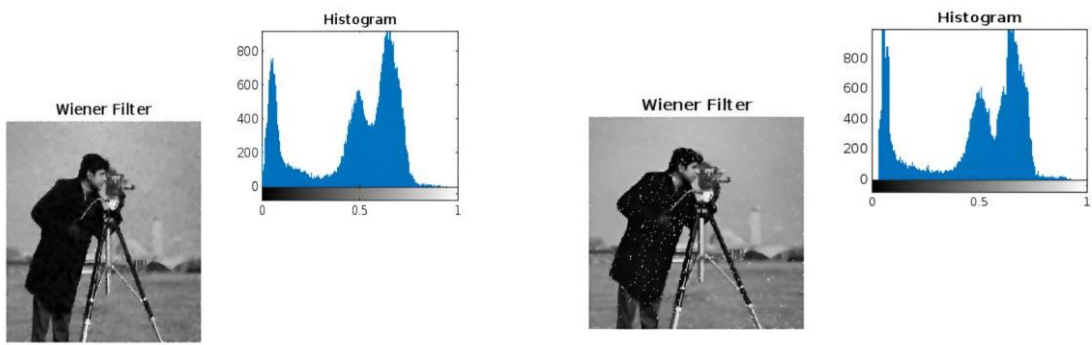

Fig. 20. Wiener Filter applied on Image with Gaussian Noise along with histogram (left side), Wiener Filter applied on Image with Salt and Pepper Noise along with histogram (right side)

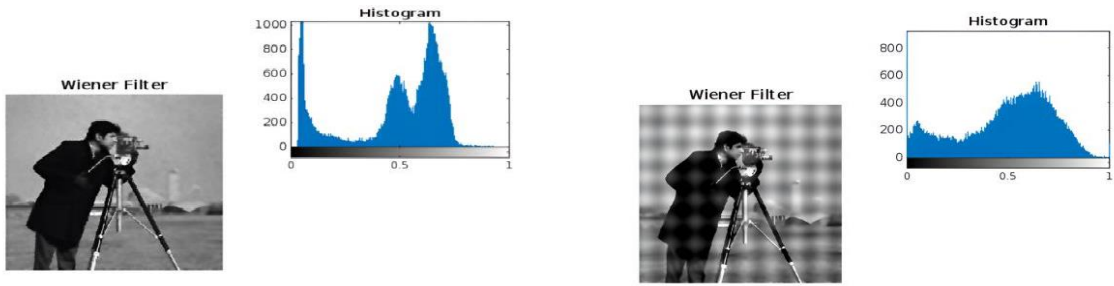

Fig. 21. Wiener Filter applied on Image with Speckle Noise along with histogram (left side), Wiener Filter applied on Image with Poisson Noise along with histogram (right side)

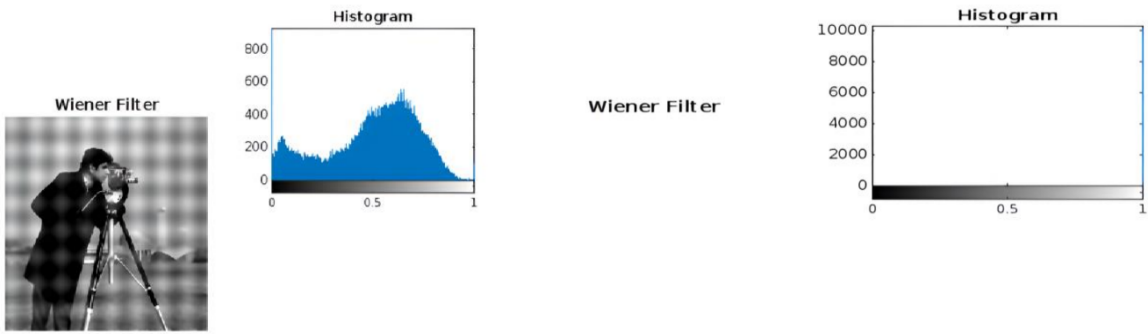

Fig. 22. Wiener Filter applied on Image with Periodic Noise along with histogram (left side), Wiener Filter applied on Image with Erlang Noise along with histogram (right side)

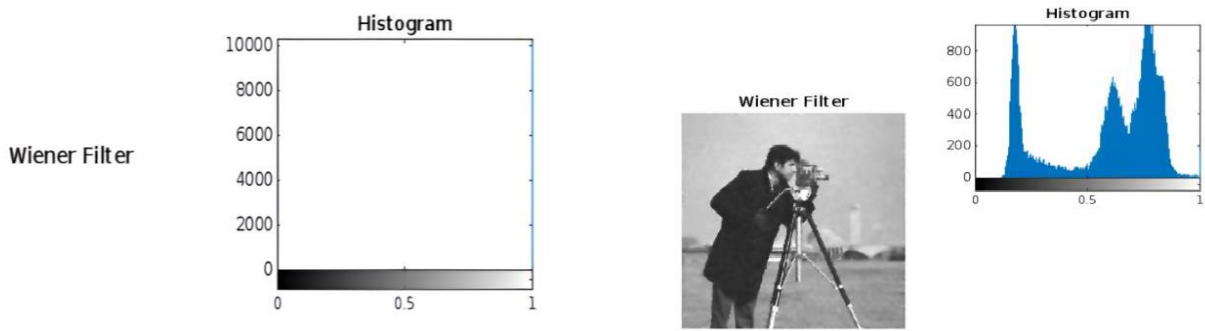

Fig. 23. Wiener Filter applied on Image with Exponential Noise along with histogram (left side), Wiener Filter applied on Image with Rayleigh Noise along with histogram (right side)

**5.4 Gaussian Filter**

A Gaussian filter is a type of linear spatial filter that is used for smoothing and noise reduction. Standard deviation is the parameter that defines the Gaussian filter. This filter works well for additive noise. It's most suited for Gaussian noise images although it can work partially effectively on other additive-type noises such as uniform noise, shot noise etc.

Gaussian noise is produced by convolving a picture with a Gaussian kernel, generating a two-dimensional matrix of Gaussian distribution values, and calculating the weighted average of local neighborhood pixel values. The Gaussian function determines the weights; greater values produce more aggressive smoothing, while smaller values produce less.
(or)
It uses a Gaussian kernel, a two-dimensional matrix, to compute a weighted average of pixel values within a local neighborhood, adjusting weights to minimize high-frequency noise and preserve key features.
The formula for Gaussian filter is

$$G(x,y) = \frac{1}{2\pi\sigma^2} e^{\frac{-(x^2+y^2)}{2\sigma^2}}$$

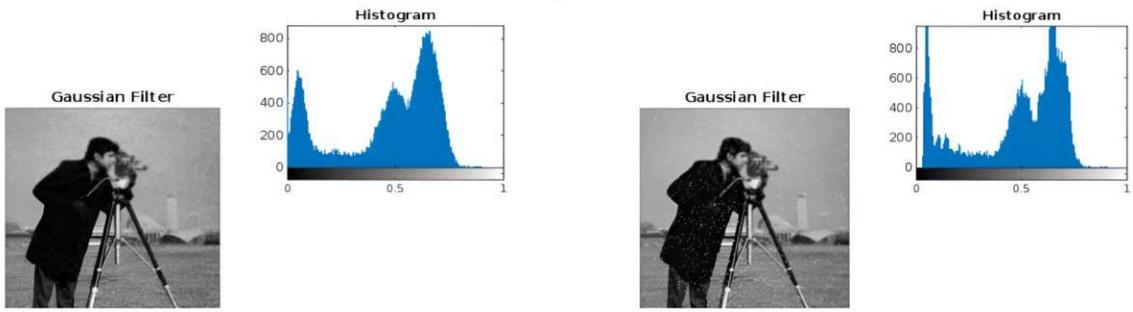

Fig. 24. Gaussian Filter applied on Image with Gaussian Noise along with histogram (left side), Gaussian Filter applied on Image with Salt and Pepper Noise along with histogram (right side)

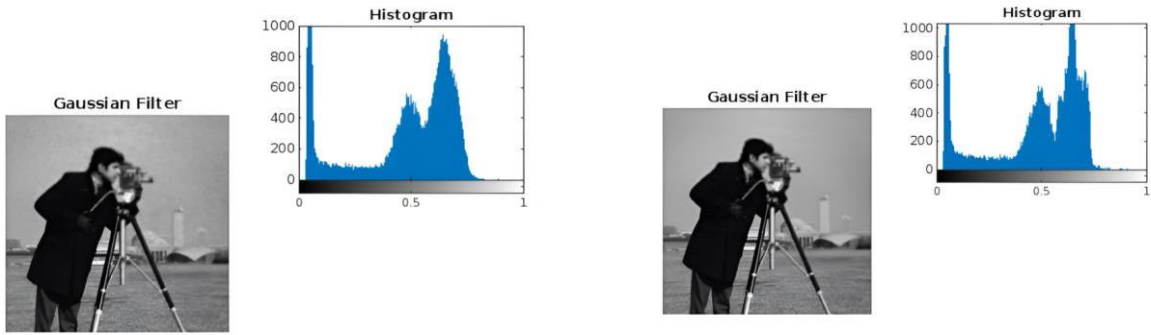

Fig. 25. Gaussian Filter applied on Image with Speckle Noise along with histogram (left side), Gaussian Filter applied on Image with Poisson Noise along with histogram (right side)

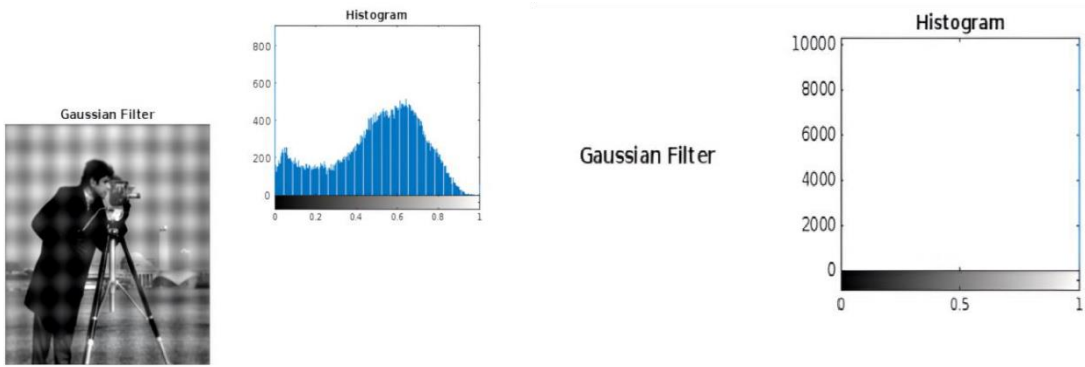

Fig. 26. Gaussian Filter applied on Image with Periodic Noise along with histogram (left side), Gaussian Filter applied on Image with Erlang Noise along with histogram (right side)

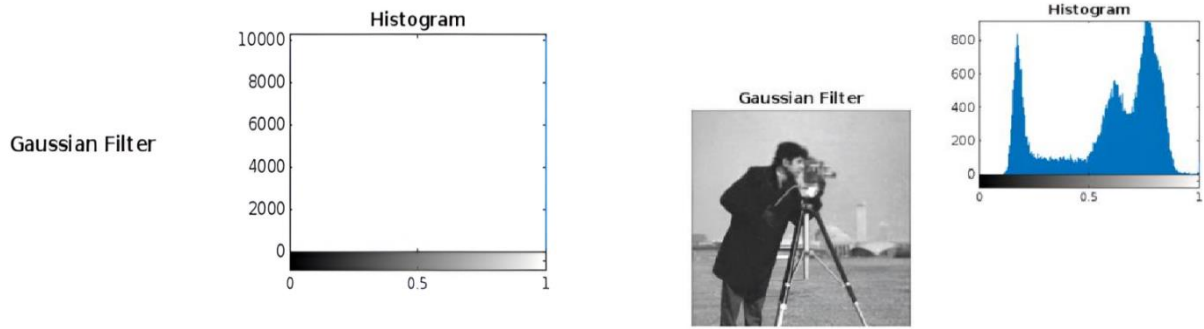

Fig. 27. Gaussian Filter applied on Image with Exponential Noise along with histogram (left side), Gaussian Filter applied on Image with Rayleigh Noise along with histogram (right side)

**5.5 Low pass Filter**

A low-pass filter is a type of linear frequency domain filter. Its objective is to eliminate high-frequency elements from an image while leaving low-frequency elements mostly unaltered. They are mostly suited for reducing the high-frequency noises in digital images. Noise models that are well-suited for low-pass filters include Gaussian Noise, Salt-and-Pepper Noise etc. It computes a weighted average of nearby and less distant pixels and the image is convolved using a filter kernel that smoothes pixel values. This preserves low-frequency features like edges and contours while suppressing high-frequency noise. A low-pass filter's efficiency is determined by its design parameters, which include the cutoff frequency and kernel size and shape.

The formula for low pass filter is

$$H(u,v) = e^{-D^2(u,v)/2\sigma^2}$$

D(u,v) is distance from the origin of Fourier Transform
Parameter : σ=$D_0$(cutoff frequency)

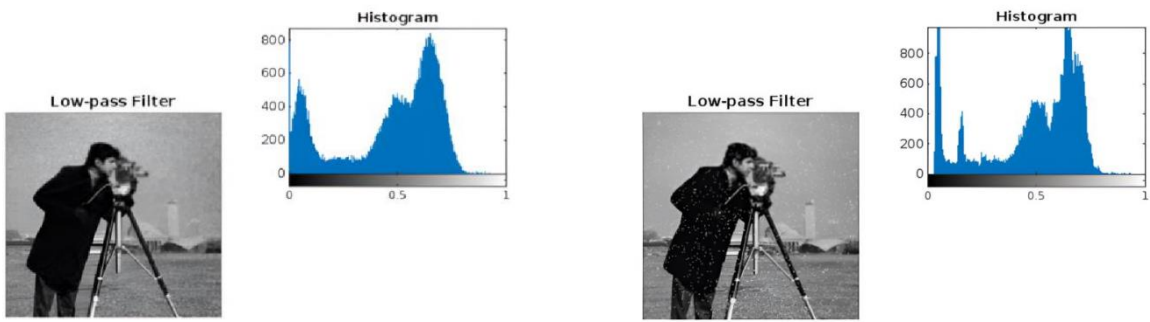

Fig. 28. Low pass Filter applied on Image with Gaussian Noise along with histogram (left side), Low pass Filter applied on Image with Salt and Pepper Noise along with histogram (right side)

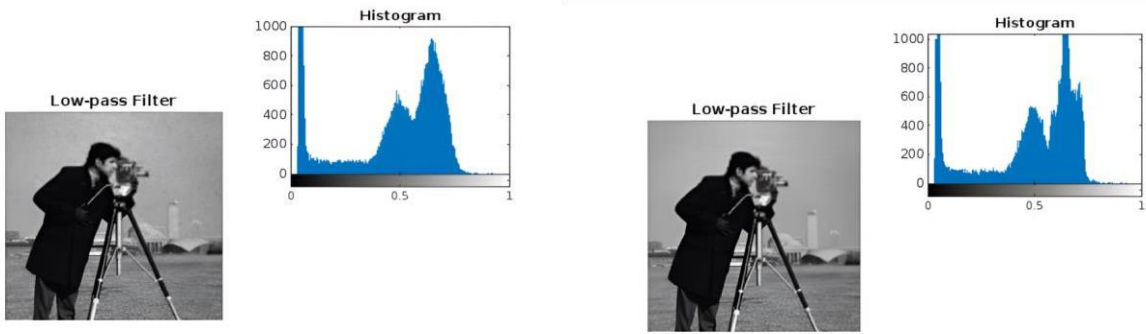

Fig. 29. Low pass Filter applied on Image with Speckle Noise along with histogram (left side), Low pass Filter applied on Image with Poisson Noise along with histogram (right side)

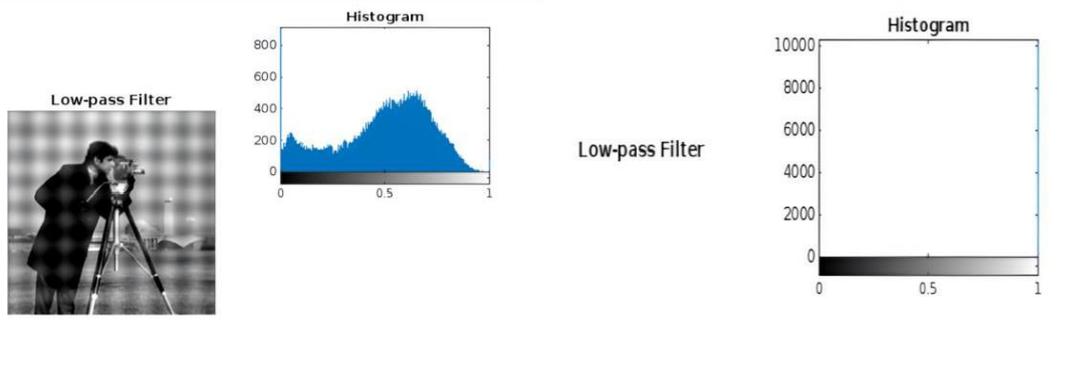

Fig. 30. Low pass Filter applied on Image with Periodic Noise along with histogram (left side), Low pass Filter applied on Image with Erlang Noise along with histogram (right side)

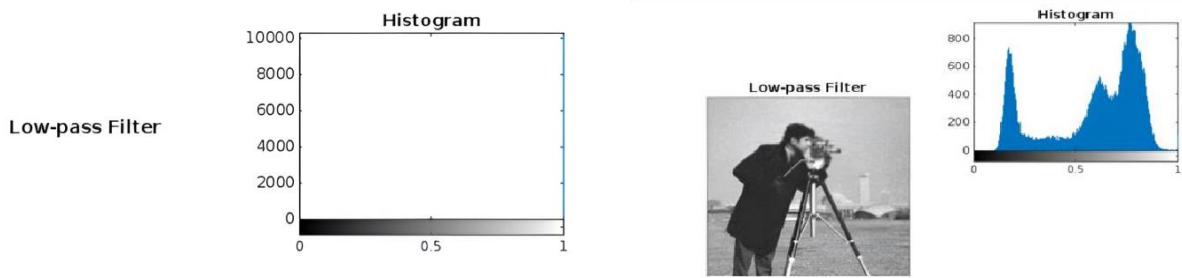

Fig. 31. Lowpass Filter applied on Image with Exponential Noise along with histogram (left side), Gaussian Filter applied on Image with Rayleigh Noise along with histogram (right side)

**5.6 High pass Filter**

High pass filter is a type of linear frequency domain filter. It is used to enhance high-frequency components while suppressing the low-frequency components. High-pass filters are used for tasks like edge detection, sharpening, and noise reduction. In the case of Gaussian noise, it lessens the impact of low-frequency Gaussian noise, such as smooth variations or gradual transitions in intensity. A high-pass filter is commonly implemented using convolution with a filter kernel in digital image processing, which amplifies high-frequency information and suppresses low-frequency information. In order to do this, the original image is subtracted from a low-pass filtered version of it. The output image that is produced has low-frequency variations, and high-frequency details that are enhanced. The formula for high pass filter is

$$H(u,v) = 1 - e^{-D^2(u,v)/2\sigma^2}$$

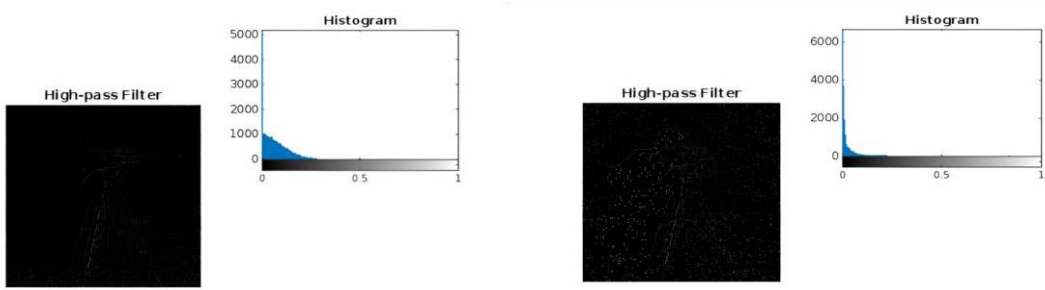

Fig. 32. High pass Filter applied on Image with Gaussian Noise along with histogram (left side), High pass Filter applied on Image with Salt and Pepper Noise along with histogram (right side)

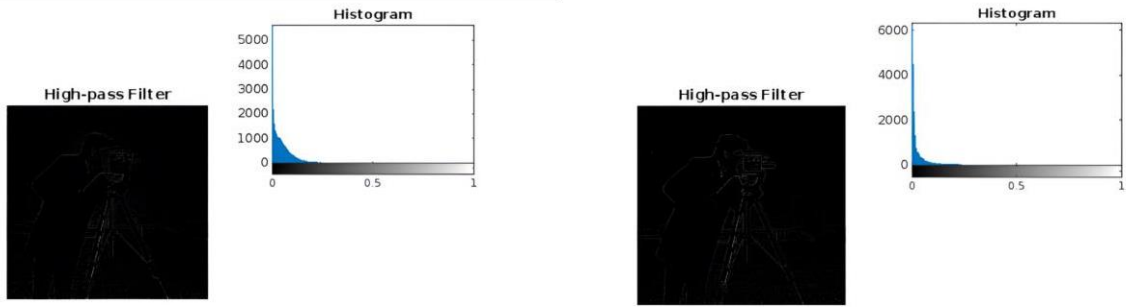

Fig. 33. High pass Filter applied on Image with Speckle Noise along with histogram (left side), High pass Filter applied on Image with Poisson Noise along with histogram (right side)

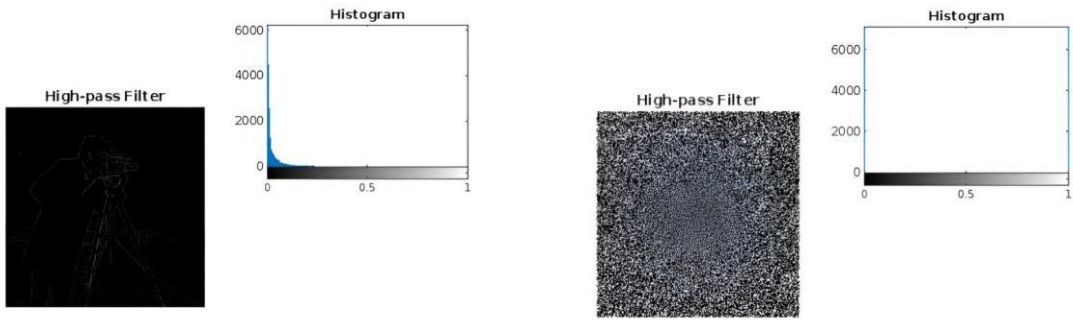

Fig. 34. High pass Filter applied on Image with Periodic Noise along with histogram (left side), High pass Filter applied on Image with Erlang Noise along with histogram (right side)

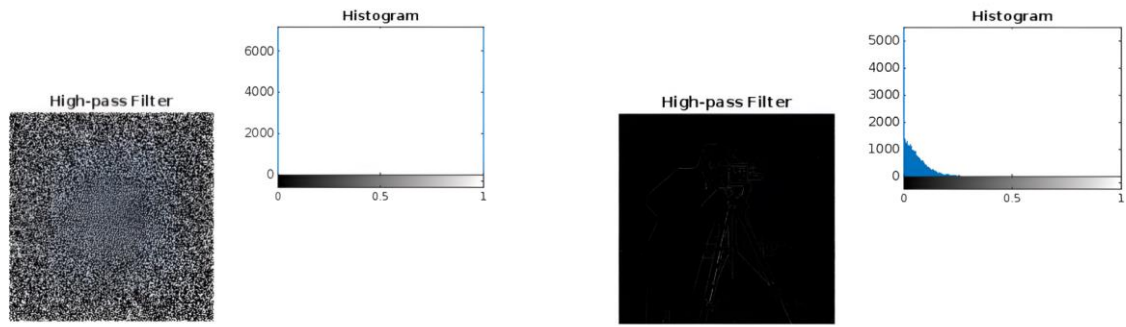

Fig. 35. High pass Filter applied on Image with Exponential Noise along with histogram (left side), High pass Filter applied on Image with Rayleigh Noise along with histogram (right side)

**5.7 Bilateral Filter**

The bilateral filter uses a non-linear filtering technique known for its capability to blur an image while simultaneously retaining sharp edges. Its unique ability to decompose the image into various scales without inducing undesirable artifacts, such as halos, post-modification, has contributed to its widespread adoption in Image Processing. The formula for bilateral filter is

$$BF[I]_p = \frac{1}{W_p} \sum_{q \in S} G_{\sigma_s}(\|p - q\|) G_{\sigma_r}(|I_p - I_q|) I_q,$$

Where normalization factor $W_p$ ensures pixel weights sum to 1.0:

$$W_p = \sum_{q \in S} G_{\sigma_s}(\|p - q\|) G_{\sigma_r}(|I_p - I_q|) I_q,$$

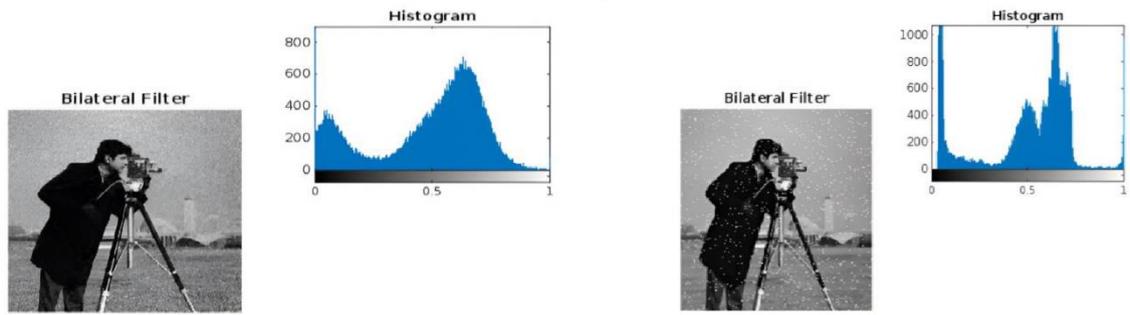

Fig. 36. Bilateral Filter applied on Image with Gaussian Noise along with histogram (left side), Bilateral Filter applied on Image with Salt and Pepper Noise along with histogram (right side)

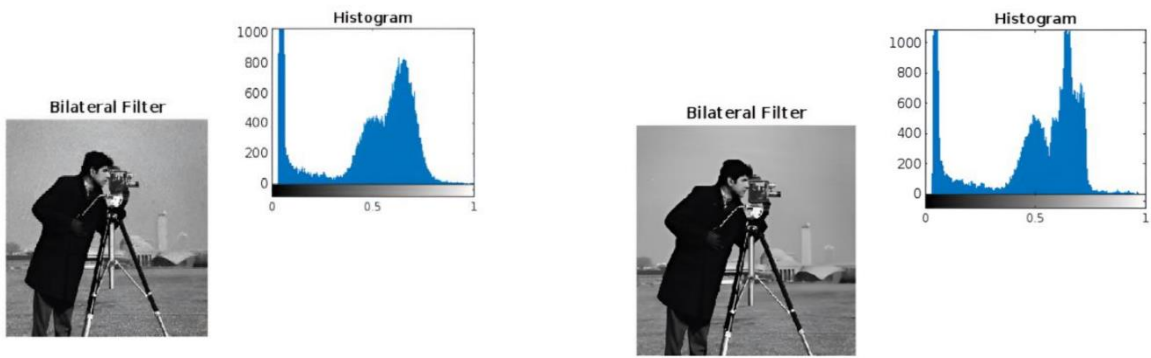

Fig. 37. Bilateral Filter applied on Image with Speckle Noise along with histogram (left side), Bilateral Filter applied on Image with Poisson Noise along with histogram (right side)

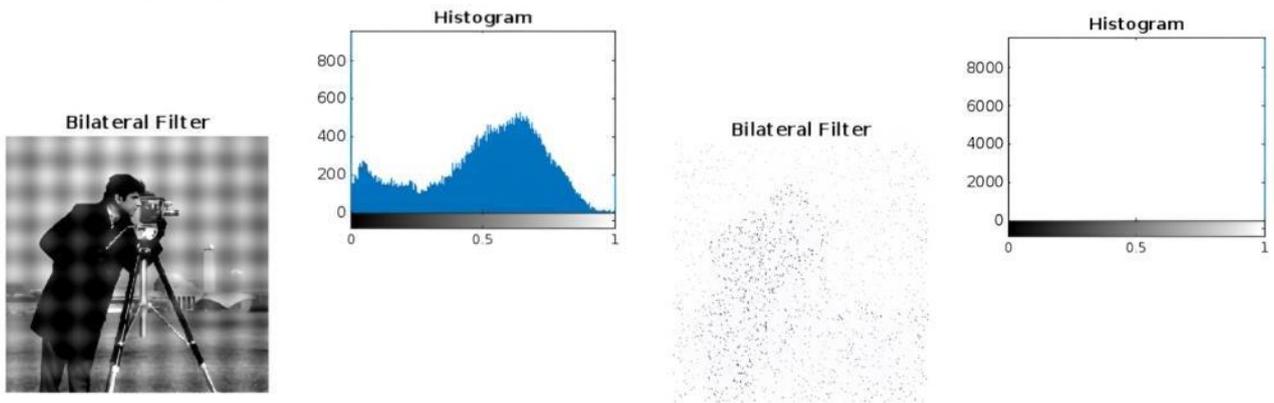

Fig. 38. Bilateral Filter applied on Image with Periodic Noise along with histogram (left side), Bilateral Filter applied on Image with Erlang Noise along with histogram (right side)

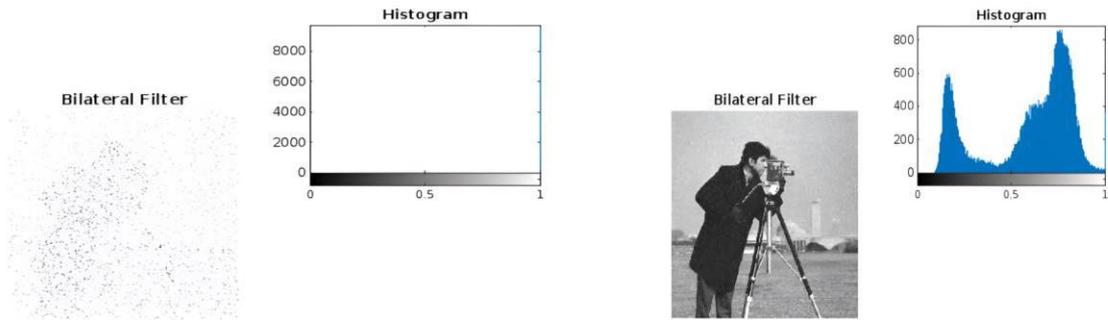

Fig. 39. Bilateral Filter applied on Image with Exponential Noise along with histogram (left side), Bilateral Filter applied on Image with Rayleigh Noise along with histogram (right side)

**5.8 Laplacian Filter**

Laplacian Filter is used to compute the second order derivative of an image to detect the edges. A laplacian filter is required for extracting the essential features of an image. In first order derivative filters we detect the horizontal direction and vertical direction edges separately and combine the both but by using the laplacian filter we detect the edges in the entire image at once.

The formula for laplacian filter is
where f denotes the image.

$$\nabla^2 f = \frac{\partial^2 f}{\partial x^2} + \frac{\partial^2 f}{\partial y^2}$$

where f denotes the image.

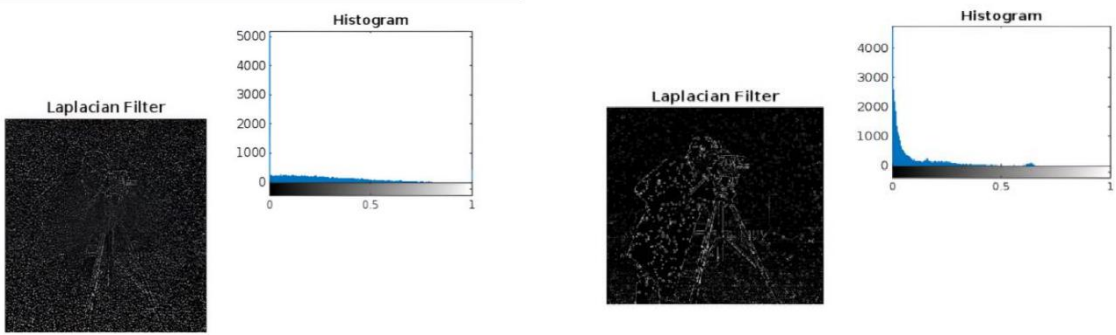

Fig. 40. Laplacian Filter applied on Image with Gaussian Noise along with histogram (left side), Laplacian Filter applied on Image with Salt and Pepper Noise along with histogram (right side)

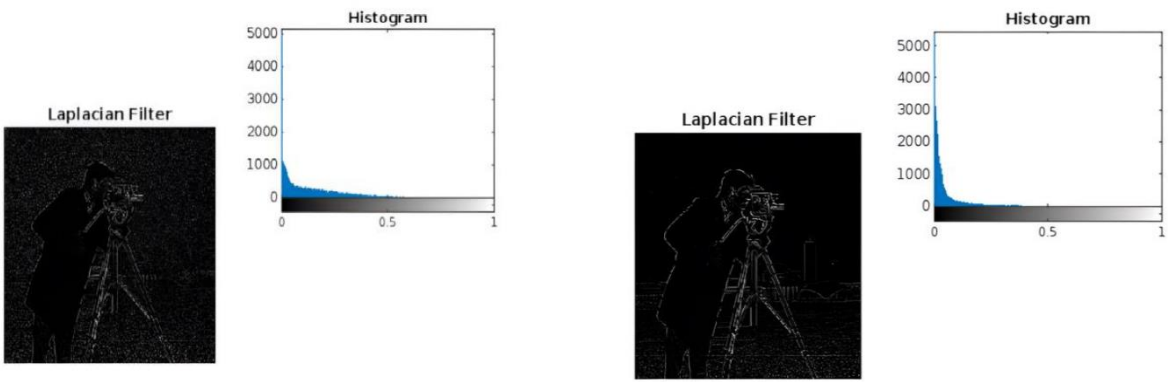

Fig. 41. Laplacian Filter applied on Image with Speckle Noise along with histogram (left side), Laplacian Filter applied on Image with Poisson Noise along with histogram (right side)

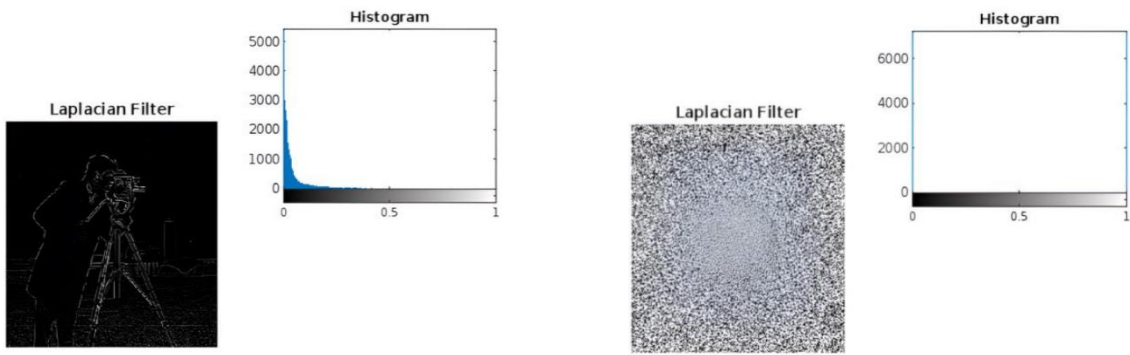

Fig. 42. Laplacian Filter applied on Image with Periodic Noise along with histogram (left side), Laplacian Filter applied on Image with Erlang Noise along with histogram (right side)

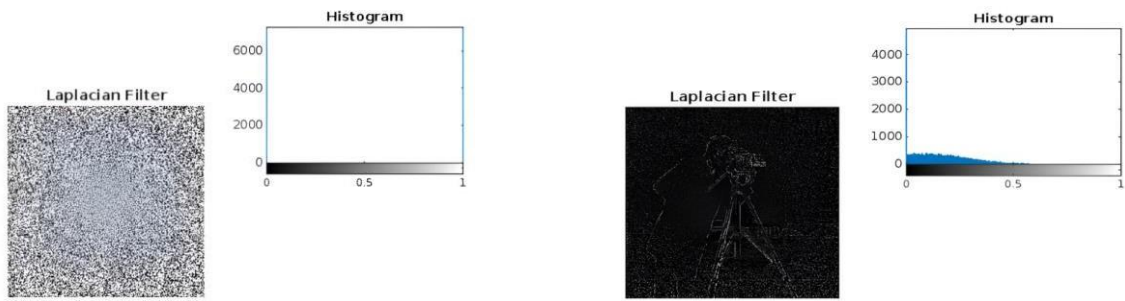

Fig. 43. Laplacian Filter applied on Image with Exponential Noise along with histogram (left side), Laplacian Filter applied on Image with Rayleigh Noise along with histogram (right side)

## 6. Conclusion and Future Research Directions

Table 1 shows the values of Peak Signal to Noise Ratio (PSNR) and highest PSNR Value of a particular noise corresponding to particular filter indicates that it is most appropriate for that particular type of noise

| PSNR Values in dB | Median Filter | Mean Filter | Wiener Filter | Gaussian Filter | Low pass Filter | High pass Filter | Bilateral Filter | Laplacian Filter |
|---|---|---|---|---|---|---|---|---|
| Gaussian Noise | 24.1406 | 23.8179 | 26.1526 | 24.3407 | 23.8179 | 5.5641 | 22.6435 | 3.2625 |
| Salt and Pepper Noise | 27.0136 | 23.7189 | 23.5102 | 24.1790 | 23.7189 | 5.5838 | 20.4714 | 3.4473 |
| Speckle Noise | 26.0616 | 24.7695 | 28.5000 | 25.1351 | 24.7695 | 5.6649 | 29.5662 | 4.2090 |
| Poisson Noise | 27.2102 | 25.1608 | 30.1726 | 25.4431 | 25.1608 | 5.7045 | 35.6945 | 4.6469 |
| Periodic Noise | 19.3043 | 18.9568 | 19.8318 | 19.0594 | 18.9568 | 5.7049 | 20.0075 | 4.642 |
| Erlang Noise | -16.4464 | -18.6268 | -18.692 | -18.5076 | -18.6268 | -17.7383 | -21.2256 | 8.7439 |
| Exponential Noise | -18.195 | -20.4282 | -20.505 | -20.3101 | -20.4282 | -19.5254 | -23.0279 | -31.1588 |
| Rayleigh Noise | 17.7283 | 17.2949 | 17.6416 | 17.3917 | 17.2949 | 5.6506 | 17.6648 | 3.9435 |

Table 1 showing PSNR Values

After evaluating all the types of filters, it is evaluated as follows

| NOISE | MOST APPROPRIATE FILTER |
|---|---|
| Gaussian Noise | Wiener Filter |
| Salt and Pepper Noise | Median Filter |
| Speckle Noise | Bilateral Filter |
| Poisson Noise | Bilateral Filter |
| Periodic Noise | Bilateral Filter |
| Erlang Noise | Laplacian Filter |
| Exponential Noise | Median Filter |
| Rayleigh Noise | Median Filter |

Table 2 represents the most appropriate filter for that particular type of noise

## Acknowledgment

We would like to express our sincere gratitude to all individuals and organizations that contributed to the success of this research. Our heartfelt thanks go to the faculties who helped in this research.

## Conflict of Interest

The authors declare no conflict of interest.

**Authors' Profiles**

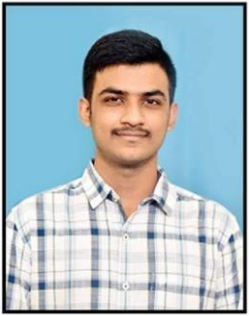

Mr. Sahil Ali Akbar is pursuing an Undergraduate Engineering degree in Computer Science and Engineering with specialization in Artificial Intelligence and Machine Learning from SRM Institute of Science and Technology. His area of interest in Artificial Intelligence, Machine Learning, Digital Image Processing, Natural Language Processing.

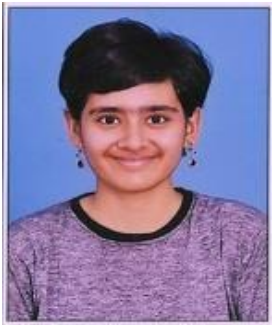

Ms. Ananya Verma is pursuing an Undergraduate Engineering degree in Computer Science and Engineering from SRM Institute of Science and Technology. Her area of interest is Quantum Computing, Digital Image Processing, Artificial Intelligence and Machine Learning.